\documentclass[a4paper]{article}
\usepackage{issp2014,amssymb,amsmath,graphicx,bm}
\usepackage[T1]{fontenc}
\usepackage[utf8]{inputenc}
\usepackage{tipa}
\usepackage{newunicodechar}
\usepackage{algorithm} 
\usepackage{algorithmic}
\usepackage[dvipsnames]{xcolor}
\usepackage{enumitem}
\usepackage{makecell}

\newunicodechar{ə}{@}
\newunicodechar{ɛ}{E}
\newunicodechar{ɪ}{I}
\newunicodechar{ʊ}{U}
\newunicodechar{ɑ}{A}
\newunicodechar{ʌ}{2}
\DeclareUnicodeCharacter{0301}{*************************************}

\definecolor{2_FDTD}{RGB}{153, 0, 76}
\definecolor{2_5FDTD}{RGB}{0, 102, 0}

\usepackage{filecontents}

\usepackage[utf8]{inputenc}
\usepackage[T1]{fontenc}
\usepackage[english]{babel} 
\usepackage[autostyle]{csquotes}
\usepackage[backend=biber,style=authoryear,sorting=nyt]{biblatex}
\defbibheading{refs}{\section{References}}
\bibliography{ref} 

\ninept

\title{A comparative study of two-dimensional vocal tract acoustic modeling based on Finite-Difference Time-Domain methods}

\makeatletter
\def\name#1{\gdef\@name{#1\\}}
\makeatother
\name{{ Debasish Ray Mohapatra$^1$, Victor Zappi$^2$, Sidney Fels$^1$}}

\address{\small \em $^1$Electrical and Computer Engineering Department, University of British Columbia, Canada\\
\small \em $^2$College of Arts, Media and Design, Northeastern University, USA\\
{\small \tt debasishray@ece.ubc.ca, v.zappi@northeastern.edu, ssfels@ece.ubc.ca}}

\begin{document}
\maketitle

\begin{abstract}

The two-dimensional (2D) numerical approaches for vocal tract (VT) modelling can afford a better balance between the low computational cost and accurate rendering of acoustic wave propagation. However, they require a high spatio-temporal resolution in the numerical scheme for a precise estimation of acoustic formants at the simulation run-time expense. We have recently proposed a new VT acoustic modelling technique, known as the 2.5D Finite-Difference Time-Domain (2.5D FDTD), which extends the existing 2D FDTD approach by adding tube depth to its acoustic wave solver. In this work, first, the simulated acoustic outputs of our new model are shown to be comparable with the 2D FDTD  and a realistic 3D FEM VT model at a low spatio-temporal resolution. Next, a radiation model is developed by including a circular baffle around the VT as head geometry. The transfer functions of the radiation model are analyzed using five different vocal tract shapes for vowel sounds /a/, /e/, /i/, /o/ and /u/.

\end{abstract}

\vspace*{1em}
\noindent \textbf{Keywords:} computational acoustics, vocal tract, FDTD, articulatory speech synthesis

\section{Introduction}

Various articulatory models have been developed to approximate the sophisticated upper vocal tract geometry and capture its acoustic features using a physics-based acoustic wave solver. The 3D acoustic analysis ~\parencite{takemoto2010acoustic,vampola2015human} can precisely characterize the vocal tract's spectral features while offering better geometrical flexibility. However, such models are not pragmatic for designing real-time speech synthesizers due to their high computational cost. In contrast to 3D, the 1D vocal tract models are computationally lightweight. Nevertheless, their over-simplified representation of a tube structure substantially eliminates its geometrical details. Hence, the acoustic analysis of 1D models is not accurate at higher formants. 

As an alternative, the 2D acoustic wave solvers ~\parencite{arnela2014two,speed2009characteristics} represent the complex vocal tract geometry as a 2D contour (a mid-sagittal cut of a 3D VT geometry) and capture the wave interaction only along the mid-sagittal plane. Due to the dimensionality reduction, the 2D models are lightweight compare to 3D ones. The most well-known 2D acoustic analysis methods for vocal tract modelling are the finite element method (FEM) and FDTD. Since 2D models do not lump off-plane wave interaction, direct use of cross-sectional areas from the vocal tract MRI images to represent 2D contours inside a simulation grid yields erroneous acoustic output. Alternatively, a nonlinear area function transformation method needs to be implemented to tune the acoustic features of 2D models that can match the more complex 3D ones. However, this procedure kills the performance boost of 2D models. Consequently, we have proposed a new vocal tract acoustic model (2.5D FDTD) that improves the 2D model by capturing the wave interaction across the mid-sagittal plane. The next section briefly discusses the numerical implementation of a 3D tube using 2D and 2.5D FDTD schemes.

\section{Methodology}
\subsection{Acoustic Wave Solver}


We implemented the FDTD numerical analysis technique to discretize the 2D acoustic wave equation on a staggered Yee grid. The 2D Yee scheme consists of a rectangular grid, define acoustic parameters (pressure ($p^{(n)}$) and velocity ($v_x^{(n)}$, $v_y^{(n)}$)) at each grid point, a squared 2D cell~\parencite{allen2015aerophones, yee1966numerical}, as shown in \textbf{Figure \ref{fig: YeeGrid_depthMap}}.  However, the employed technique for acoustic simulation does not facilitate dynamic boundary conditions. Hence, a new scalar field $0 \leq \beta(x, y, t) \leq 1$ was introduced to the solver, which could transit smoothly between $\beta=1$ (air) and $\beta=0$ (boundary). At $\beta=0$, a prescribed velocity $v=\bm{v_{b}}$ was enforced to handle the boundary condition. Specifically, per each time step ($n$),  the wave solver updates pressure component $p^{(n)}$ and velocity components ($v_x^{(n)}$, $v_y^{(n)}$) at each grid point, by solving the below linear wave equations in the time domain~\parencite{zappi2016towards}:

\vspace{-10pt}
\begin{align}
\frac{\partial p}{\partial t} +(1 - \beta)p \ = \ -  \rho c^2 (  \ \frac{\partial v_x }{\partial x}  \ + \ \frac{\partial v_y }{\partial y} \ ) 
\label{eq:wave_cont1}\\
\beta\frac{\partial\bm{v}}{\partial t} \ + \ \left(1-\beta\right)\bm{v} \ = \ -\beta^{2}\frac{\bm{\nabla} p}{\rho} \ + \ \left(1-\beta\right)\bm{v}_{b}
\label{eq:wave_mot1}
\end{align}
\vspace{-7pt}

The 2.5D FDTD follows the above rationale but improves upon it by incorporating new impedance terms in its 2D acoustic wave solver, known as \textit{tube depth}.  The tube's depth $(D)$, i.e., its continuous extension along the $z$ axis (with $x$ and $y$ axis being the dimension of starting 2D scheme), is derived from the cross-sectional area of the tube and sampled at each point of the scheme as shown in \textbf{Figure \ref{fig: YeeGrid_depthMap}}. Then the resulting depths are mapped to their respective acoustic parameters ($p$, $\bm{v_{x}}$, $\bm{v_{y}}$) as demonstrated by~\textcite{mohapatra2019extended}. The inclusion of tube depth allows the wave solver to apprehend wave interactions across the mid-sagittal plane, in turn eliminating the nonlinear cross-sectional area functions transformation requirement. Therefore, it is expected that the time complexity of a 2.5D FDTD tube model will be better than the existing 2D models. Basically, with the help of following discrete update rules, we adopt a time marching algorithm (see \textbf{Algorithm~\ref{algo:fdtd_timemarching}}) to sample the acoustic parameters at every grid point:

\begin{figure}[!th]
\includegraphics[width=\columnwidth]{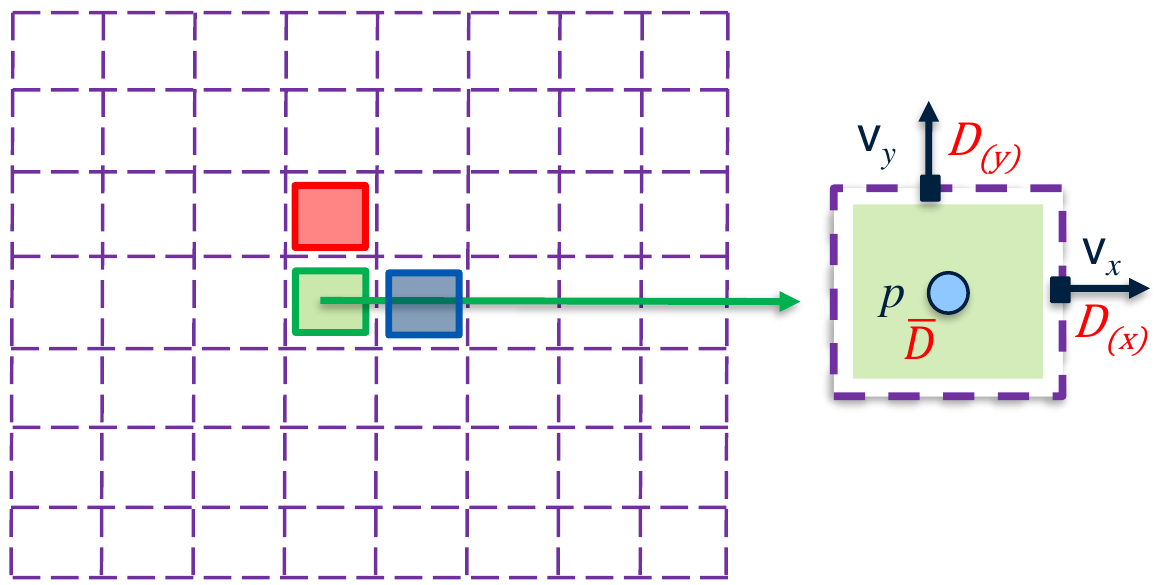}
\vspace{-15pt}
\caption{Representation of acoustic parameters ($p$, $\bm{v_{x}}$, $\bm{v_{y}}$) and tube depth ($\bar{D}$, $D_{(x)}$ and $D_{(y)}$) inside the 2D Yee scheme }
\label{fig: YeeGrid_depthMap}
\vspace{-25pt}
\end{figure}

\vspace{-10pt}
\begin{align}
& p^{(n+1)} \ = \ \frac{ \bar{D} p^{(n)}  - \rho c^2 \Delta  t \ \bm{\widetilde{\nabla}} \cdot \bm{V}^{(n)} } { \bar{D}} 
\label{eq:fdtd_cont}\\
&\bm{v}^{(n+1)} \ = \ \frac{ \beta \bm{v}^{(n)}  - \beta^2 \Delta  t \ \widetilde{\nabla} p^{(n+1)} / \rho \ + \  \Delta t(1-\beta) \bm{v_b}} { \beta + \Delta t(1-\beta)} 
\label{eq:fdtd_mot}
\end{align}
\vspace{-17pt}

with:

\vspace{-17pt}
\begin{flalign}
& \bm{V} \ = \ ( \ D_{(x)} v_x,  \ D_{(y)} v_y \ )
\label{eq:extended_vel}
\end{flalign}
\vspace{-10pt}

\begin{algorithm}[!h]
        \algsetup{linenosize=\scriptsize}
        \scriptsize
        \caption{FDTD Time-Marching algorithm}
        \hspace*{\algorithmicindent} \textbf{Input:} VT area function $a(x)$, audio time $t$\\
        \begin{algorithmic}[1]
            \STATE Initialize physical constants: air density $\rho$, sound speed $c$, boundary admittance coefficient $\mu$.
            
            \STATE Initialize the simulation sampling rate $R$.
            
            \STATE Set the temporal resolution ($\Delta t$) and grid resolution ($\Delta s$) with $R$ and CFL condition.
            
            \STATE Create a optimized grid having size $(M\times N)$.
            
            \STATE Set the total number of time steps $T$ of the simulator using $t$ and $\Delta t$.
            
            \STATE Define boundary cells with $a(x)$  and normal acoustic impedance $Z$.

            \STATE Set depth values ($\bar D$, $D_{(x)}$ and $D_{(y)}$) for each grid cell from $a(x)$.
            
            \STATE Define source excitation cells
                        
            \STATE Initialize source excitation velocity $v_e$.
            
            \STATE Initialize acoustic components ($p$, $v_x$ and $v_y$) for each grid cell.
            
            \FOR{$n=1...T$}
                \FOR{$i=1...M$}
                    \FOR{$j=1...N$}
                    \STATE Update $p^{n+1}(i,j)$ with $v^n_x(i,j)$, $v^n_y(i,j)$ and $D(i,j)$ (Eq. \ref{eq:fdtd_cont} and Eq.\ref{eq:extended_vel})
                        \IF{$(i,j)=$ excitation cell}
                            \STATE $v^n_x \leftarrow v^n_x+v^n_e$ and $v^n_y \leftarrow v^n_y+v^n_e$
                        \ENDIF
                        \IF{$(i,j)=$ boundary cell}
                            \STATE $v^n_b \leftarrow p^n(i,j)/Z$
                        \ENDIF
                        \STATE Update $v^{n+1}_x(i,j)$ with $p^{n+1}(i,j)$ and $v^n_b$ (Eq. \ref{eq:fdtd_mot})
                        \STATE Update $v^{n+1}_y(i,j)$ with $p^{n+1}(i,j)$ and $v^n_b$ (Eq. \ref{eq:fdtd_mot})
                    \ENDFOR
                \ENDFOR
            \ENDFOR
        \end{algorithmic}
        \label{algo:fdtd_timemarching}
\end{algorithm}
\vspace{-11pt}

\subsection{Vocal Tract Wall Losses}
To simulate vocal tract wall reflection, we adopted the local reactive boundary approach as proposed here~\parencite{yokota2002visualization,takemoto2010acoustic}. The method estimates wall losses (locally reacting soft walls) and enforces boundary condition by using $\bm{v_w}$, the normal particle velocity going into or coming from a wall.
\begin{equation}
    \bm{v_w} = \frac{p_w}{Z_n} \bm{\hat{n}}
    \label{eq: boundary condition}
\end{equation}
where $p_w$ is the current pressure value in an air cell, located in front of the wall cell. And  $Z_n$, the normal acoustic impedance can be computed by using the normal sound absorption coefficient $\alpha_n$ as follows,

\begin{equation}
    Z_n = \rho c \frac{1+\sqrt{1-\alpha_n}}{1-\sqrt{1-\alpha_n}}
    \label{eq: boundary_impedance}
\end{equation}

The boundary absorption coefficient $\alpha_n$ can be derived from the boundary admittance coefficient $\mu \in (0,1)$. During the simulation, the solver combines equation (\ref{eq: boundary condition}) with equation (\ref{eq:fdtd_mot}) by setting $\bm{v_b}=\bm{v_w}$ for all the velocity vectors which are positioned in between an air cell and wall cell. However, this method does not comprehend frequency dependent wall losses.

\subsection{Excitation Model}

For acoustic analysis and speech production, articulatory models require a source excitation function just above the glottal-end to input acoustic energy into the tube~\parencite{mohapatra2018limitations}. In this work, we are only characterizing the acoustic behaviour of a vocal tract tube through its transfer function analysis. Hence, a band-passed velocity pulse having frequency range 2Hz-20kHz was injected near the glottis for various vocal tract shapes, and the corresponding transfer functions (impulse response) were obtained at the mouth-end.

\subsection{Radiation Model}

\begin{figure}[!t]
\includegraphics[width=\columnwidth]{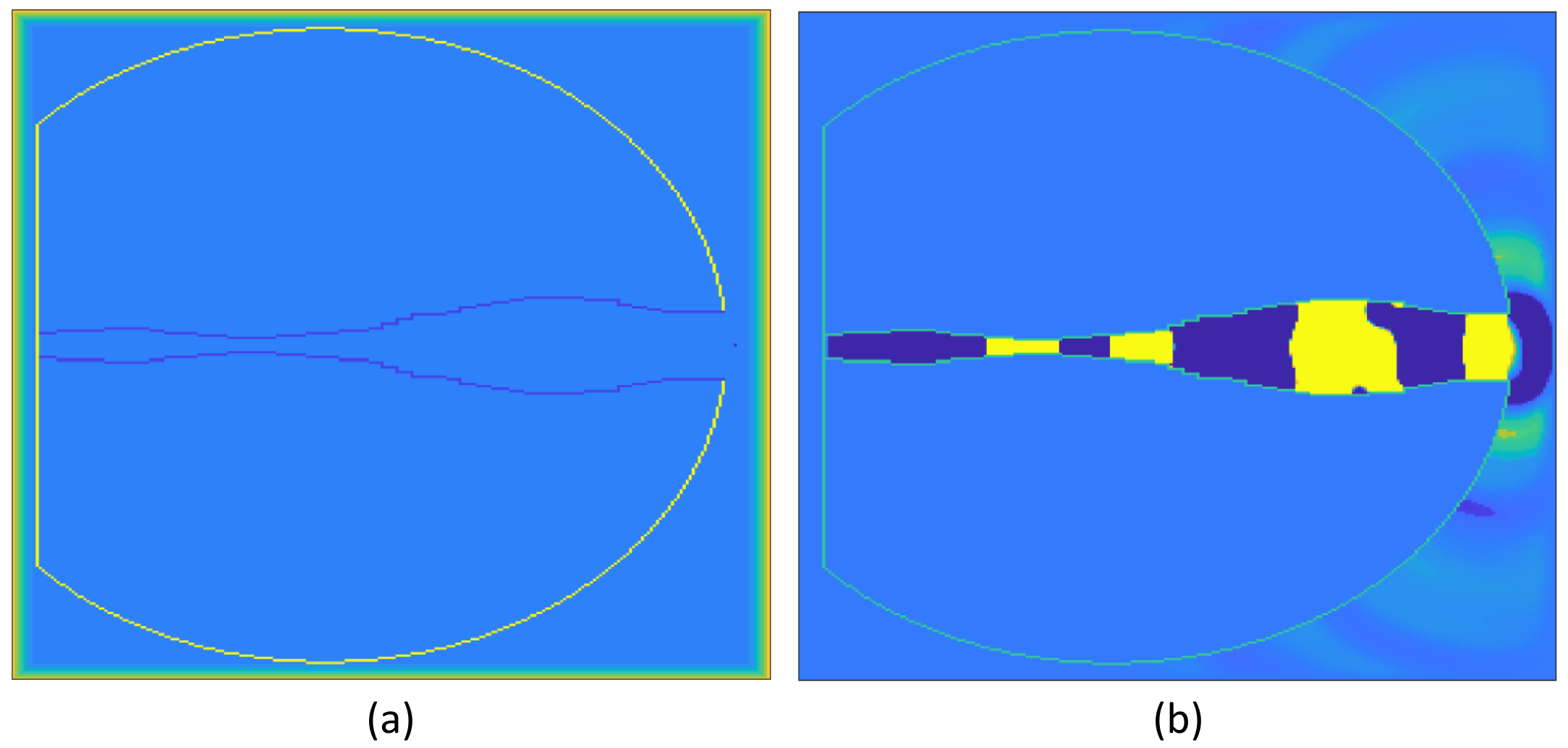}
\vspace{-15pt}
\caption{(a) Simplified head geometry representation with circular mouth aperture (b) Snapshot of acoustic wave propagation inside the vocal tract tube and emanating from mouth aperture at time instant $t=6.1ms$ for vowel /a/.}
\label{fig: RadiationModel}
\vspace{-20pt}
\end{figure}

The outward wave propagation at the mouth-end (lips) contributes to acoustic energy losses in the vocal tract. It is one of the essential dissipation mechanisms in voice production. Hence, a nonzero radiation model of a vocal tract influences the synthesized audio output by lowering the formant frequencies and increasing their bandwidths. It might be challenging to build a model that can precisely attain the radiation effects, considering a realistic human head geometry. However, it has already been demonstrated that the inclusion of a spherical baffle (a simplified approximation of the actual head geometry) around the 3D vocal tract model offers promising results for vowel production~\parencite{arnela2013effects}. Since the 2D model uses a mid-sagittal cut of a 3D vocal tract, the spherical baffle can be replaced by a circular baffle for the 2.5D radiation model as shown in \textbf{Figure~\ref{fig: RadiationModel}(a)}. Moreover, to illustrate acoustic waves' propagation emanating from the mouth toward infinity (\textbf{Figure~\ref{fig: RadiationModel}(b)}), the computational domain can be extended out of the vocal tract by using perfectly matched layers (PMLs). PMLs help in eliminating spurious wave reflection at the domain boundaries.

We create a fixed-sized circular baffle with a diameter of 0.20m around the 2.5D vocal tract contour inside the simulation domain. The circular baffle center has been adjusted to fit in the entire vocal tract contour while connecting its exit cross-section. Inside the 2.5D scheme, the region enclosed by the vocal tract and baffle boundary is set to air cells. In order to absorb outgoing waves and approximate infinite space,  $6$ layers of PML are added at all sides of the domain boundary.

\section{Experimental Setup}

\begin{figure}[!t]
\includegraphics[width=\columnwidth]{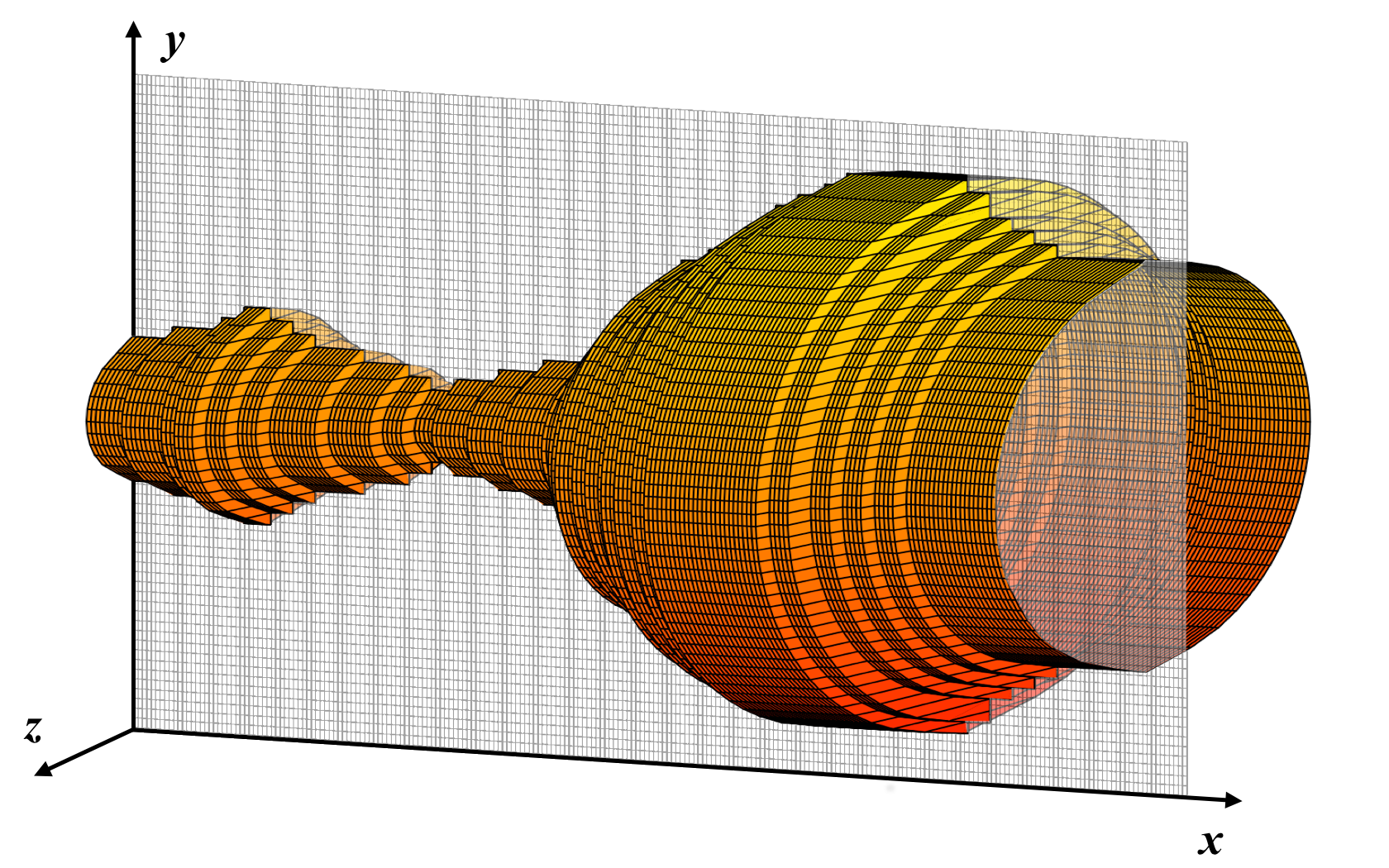}
\vspace{-20pt}
\caption{2.5D vocal tract tube representation for vowel /a/. }
\label{fig:2_5Dgeometry}
\vspace{-17pt}
\end{figure}

A comparative study was carried out for two different scenarios to characterize the acoustic features of the 2.5D vocal tract model: (1) numerical simulation with computational domain ended at the mouth aperture by imposing Dirichlet boundary conditions (open-end boundary condition) and (2) numerical simulation with free-field radiation, which includes head geometry and PML layers (radiation condition). These conditions were examined using the transfer functions analysis approach for five different vocal tract shapes(vowel sounds: /a/, /e/, /i/, /o/ and /u/). The 2.5D vocal tract shapes were derived using Story's area function dataset~\parencite{story2008comparison}, as it is the standard one for the vocal tract acoustic analysis. We have only considered the vocal tract geometry as a straight tube with circular cross-sections for this study as shown in \textbf{Figure~\ref{fig:2_5Dgeometry}}.

For open-end boundary conditions, we use a precise 3D FEM vocal tract model as the reference for the transfer function analysis. A virtual microphone is placed 3mm inside the mouth opening to record pressure variations. In the case of free-field radiation, we compare the transfer functions of the 2.5D model with Story's calculated formants, obtained from the frequency-domain analysis of vocal tracts~\parencite{sondhi1987hybrid}. And the microphone is positioned 3mm outside of the mouth to capture the radiating acoustic waves. Courant-Friedrichs-Lewy (CFL) stability condition in two dimensions: $\Delta t \leq \Delta s/\sqrt{2}c$ is imposed, where $c$ is the speed of sound. 

The simulation generates $50$ms of synthesized audio for every vowel sounds. During the simulation, we set the following physical parameters fixed: air density $\rho=1.14$ kg/m\textsuperscript{3}, boundary admittance $\mu=0.005$ and sound speed $c=350$ m/s.  We implement both the 2D and 2.5D vocal tract models in the MATLAB environment. The simulation runs on a workstation equipped with an Intel Core i7-8700K processor. The custom code for the 2.5D vocal tract model is publicly available here\footnote{\label{fn:data_repo}\textcolor{blue}{https://github.com/Debasishray19/Talking-Tube}}.

\section{Model Validation}

We follow the below steps to analyze the acoustic behaviour of the 2.5D model:
\begin{enumerate} [label= \textbf{(\Alph*)}]
	\item Transfer function analysis with open-end boundary condition for the 2D FDTD and 2.5D FDTD vocal tract models (vowel sound: /a/, /i/ and /u/).
	\item Model performance evaluation (simulation run-rime).
	\item Transfer function analysis for the 2.5D vocal tract model with free-field radiation (vowel sound: /a/, /e/, /i/. /o/ and /u/).
\end{enumerate}

For Step (A) and (B), we simulate each vowel sound under three different spatial grid resolution for both 2D and 2.5D FDTD methods: \textit{low} ($\Delta s = 0.74$mm), \textit{mid} ($\Delta s = 0.28$mm) and \textit{high} ($\Delta s = 0.18$mm). The high spatial resolution should produce precise acoustic output as it yields vocal tract geometry with greater details. However, it requires a larger simulation domain size. The grid resolution values are empirically derived. Step (A) and (B) examine the improvement in a 2.5D vocal tract model over its 2D simulation due to the addition of tube depth. Step (C) investigates the change in transfer function with the inclusion of a circular baffle for the radiation model.

\section{Result}
We applied Fast Fourier Transformation (FFT) to the recorded pressure waves to obtain their transfer function. Formants were extracted by considering local maxima in transfer function curves. Here we are only analyzing the first three formants since they are mainly responsible for distinguishing vowel sounds~\parencite{vampola2015human}. The formants for the 3D FEM model can be found here~\parencite{thesis2015Arnela}.

\subsection{Step A: Transfer Function Analysis}

Transfer function analysis for the 2D and 2.5D simulation with open-end boundary condition\footnote{\label{fn:color_map}Colormap: \textcolor{2_FDTD}{For 2D simulation}, \textcolor{2_5FDTD}{For 2.5D simulation}.}\textsuperscript{,}\footnote{All the formant frequencies are in Hz.}.

\vspace{-5pt}
\begin{table}[!ht]
    \centering
    \begin{tabular}{||c|c|c|c||}
    \hline 
     Resolution ($\Delta s$) & F1 &F2 & F3 \\
     \hline \hline
     low (\% Error)  & \makecell{\textcolor{2_FDTD}{660(-5.16)}\\\textcolor{2_5FDTD}{700(0.57)}} & \makecell{\textcolor{2_FDTD}{1040(-2.61)}\\\textcolor{2_5FDTD}{1040(-2.61)}} & \makecell{\textcolor{2_FDTD}{2960(-2.33)}\\\textcolor{2_5FDTD}{3020(-0.36)}}\\ [0.5ex]
     \hline
	 mid (\% Error) & 
	 \makecell{\textcolor{2_FDTD}{660(-5.16)}\\\textcolor{2_5FDTD}{680(-2.29)}} & \makecell{\textcolor{2_FDTD}{1040(-2.61)}\\\textcolor{2_5FDTD}{1060(-0.74)}} & \makecell{\textcolor{2_FDTD}{3000(-1.02)}\\\textcolor{2_5FDTD}{3060(0.95)}}\\ [0.5ex]
	 \hline
	 high(\% Error) & 
	 \makecell{\textcolor{2_FDTD}{700(0.57)}\\\textcolor{2_5FDTD}{680(-2.29)}} & \makecell{\textcolor{2_FDTD}{1060(-0.74)}\\\textcolor{2_5FDTD}{1040(-2.62)}} & \makecell{\textcolor{2_FDTD}{3020(-0.36)}\\\textcolor{2_5FDTD}{3040(0.29)}}\\ [0.5ex]
	 \hline
    \end{tabular}
    \caption[optionalArg]{Positional errors of the first three formants computed for vowel sound /a/, with respect to 3D FEM model.}
    \label{tab: stepA_vowel_a}
\vspace{-15pt}
\end{table}

\begin{table}[!ht]
    \centering
    \begin{tabular}{||c|c|c|c||}
    \hline 
     Resolution ($\Delta s$) & F1 &F2 & F3 \\
     \hline \hline
     low (\% Error)  & \makecell{\textcolor{2_FDTD}{240(-8.74)}\\\textcolor{2_5FDTD}{260(-1.14)}} & \makecell{\textcolor{2_FDTD}{2120(0.42)}\\\textcolor{2_5FDTD}{2160(2.32)}} & \makecell{\textcolor{2_FDTD}{3020(0.33)}\\\textcolor{2_5FDTD}{3060(1.66)}}\\ [0.5ex]
     \hline
	 mid (\% Error) & 
	 \makecell{\textcolor{2_FDTD}{260(-1.14)}\\\textcolor{2_5FDTD}{260(-1.14)}} & \makecell{\textcolor{2_FDTD}{2140(1.37)}\\\textcolor{2_5FDTD}{2140(1.37)}} & \makecell{\textcolor{2_FDTD}{3000(-0.33)}\\\textcolor{2_5FDTD}{3040(0.99)}}\\ [0.5ex]
	 \hline
	 high(\% Error) & 
	 \makecell{\textcolor{2_FDTD}{260(-1.14)}\\\textcolor{2_5FDTD}{260(-1.40)}} & \makecell{\textcolor{2_FDTD}{2140(1.37)}\\\textcolor{2_5FDTD}{2160(2.32)}} & \makecell{\textcolor{2_FDTD}{3020(0.33)}\\\textcolor{2_5FDTD}{3040(0.99)}}\\ [0.5ex]
	 \hline
    \end{tabular}
    \caption[optionalArg]{Positional errors of the first three formants computed for vowel sound /i/, with respect to 3D FEM model.}
    \label{tab: stepA_vowel_i}
\vspace{-15pt}
\end{table}

\begin{table}[!ht]
    \centering
    \begin{tabular}{||c|c|c|c||}
    \hline 
     Resolution ($\Delta s$) & F1 &F2 & F3 \\
     \hline \hline
     low (\% Error)  & \makecell{\textcolor{2_FDTD}{260(0.38)}\\\textcolor{2_5FDTD}{260(0.38)}} & \makecell{\textcolor{2_FDTD}{700(-7.52)}\\\textcolor{2_5FDTD}{720(-4.88)}} & \makecell{\textcolor{2_FDTD}{2260(-0.17)}\\\textcolor{2_5FDTD}{2300(1.59)}}\\ [0.5ex]
     \hline
	 mid (\% Error) & 
	 \makecell{\textcolor{2_FDTD}{220(-15.0)}\\\textcolor{2_5FDTD}{260(0.38)}} & \makecell{\textcolor{2_FDTD}{600(-20.7)}\\\textcolor{2_5FDTD}{700(-7.52)}} & \makecell{\textcolor{2_FDTD}{2260(-0.17)}\\\textcolor{2_5FDTD}{2300(1.59)}}\\ [0.5ex]
	 \hline
	 high(\% Error) & 
	 \makecell{\textcolor{2_FDTD}{260(0.38)}\\\textcolor{2_5FDTD}{260(0.38)}} & \makecell{\textcolor{2_FDTD}{700(-7.52)}\\\textcolor{2_5FDTD}{720(-4.88)}} & \makecell{\textcolor{2_FDTD}{2280(0.70)}\\\textcolor{2_5FDTD}{2300(1.59)}}\\ [0.5ex]
	 \hline
    \end{tabular}
    \caption[optionalArg]{Positional errors of the first three formants computed for vowel sound /u/, with respect to 3D FEM model.}
    \label{tab: stepA_vowel_u}
\vspace{-15pt}
\end{table}

\subsection{Step B: Model Performance}

The computational cost of FDTD vocal tract model depends upon the spatial grid resolution and simulation domain size. We implement an algorithm that enforces an optimized domain size for FDTD models. The table below presents the simulation domain size and time duration corresponding to each grid resolution for vowel /u/. 

\begin{table}[!ht]
    \centering
    \begin{tabular}{||c|c|c||}
    \hline 
     Resolution ($\Delta s$) & \makecell{Domain Size\\ (\textit{width$\times$height})} & \makecell{Duration (Approx.)\\(\textit{in seconds})}\\
     \hline \hline
     low ($0.74 mm$) & \makecell{$266\times 35$} & \makecell{280}\\ [0.5ex]
     \hline
	 mid ($0.28 mm$) & \makecell{$702\times 85$} & \makecell{$4.98\text{e+}3$}\\ [0.5ex]
	 \hline
	 high($0.18 mm$) & \makecell{$1051\times127$} & \makecell{$1.85\text{e+}4$}\\ [0.5ex]
	 \hline
    \end{tabular}
    \caption[optionalArg]{Simulation run-time at different grid resolutions.}
    \label{tab: stepB_model_performance}
\vspace{-15pt}
\end{table}

\subsection{Step C: Transfer Functions Analysis}

The table below presents transfer function analysis for the 2.5D vocal tract models with the free-field radiation. We have only considered the low grid resolution ($\Delta s = 0.74$mm) for simulation of vocal tract radiation models.

\begin{table}[!ht]
    \centering
    \begin{tabular}{||c|c|c|c||}
    \hline 
     Vowel & \makecell{F1}& \makecell{F2} & \makecell{F3}\\
     \hline \hline
     /a/ & \makecell{$640$\\$-7.78\%$} & \makecell{$1020$\\$8.28\%$} & \makecell{$3000$\\$1.76\%$}\\ [0.5ex]
     \hline
	 /e/ & \makecell{$340$\\$-29.75\%$} & \makecell{$1920$\\$1.74\%$} & \makecell{$2320$\\$-3.49\%$}\\ [0.5ex]
	 \hline
	 /i/ & \makecell{$260$\\$-20.00\%$} & \makecell{$2080$\\$-2.75\%$} & \makecell{$2960$\\$-0.53\%$}\\ [0.5ex]
	 \hline
	 /o/ & \makecell{$480$\\$-3.80\%$} & \makecell{$760$\\$-1.04\%$} & \makecell{$2340$\\$-3.34\%$}\\ [0.5ex]
	 \hline
	 /u/ & \makecell{$260$\\$-17.19\%$} & \makecell{$720$\\$2.56\%$} & \makecell{$2180$\\$-5.13\%$}\\ [0.5ex]
	 \hline
    \end{tabular}
    \caption[optionalArg]{Formant frequencies and the percentage positional errors of the first three formants for the vocal tract radiation model.}
    \label{tab: stepC_TransferFunction_RadiationModel}
\vspace{-15pt}
\end{table}

\section{Discussion \& Conclusion}

The transfer function analysis is a standard practice to validate acoustic features of area-based articulatory models. In the first step of our evaluation, the 2.5D FDTD simulation produces positional errors below $2\%$ at a low grid resolution. Moreover, this characteristic is consistent across all three vowels. However, the positional error is high for the 2D FDTD simulation at a low grid resolution in most cases. \textbf{Table~\ref{tab: stepA_vowel_a}, \ref{tab: stepA_vowel_i} and \ref{tab: stepA_vowel_u}} show that the acoustic outputs of the 2D FDTD model can be tuned by increasing its computational grid resolution whereas this approach doesn't make much difference for the 2.5D FDTD simulation. \textbf{Table~\ref{tab: stepB_model_performance}} shows for a real-time speech synthesizer the low grid resolution is always desirable.

In the case of free-field of radiation, \textbf{Table~\ref{tab: stepC_TransferFunction_RadiationModel}} presents the first three formants and their positional errors. The positional error of the first formant is high across all the vowels. And the second and third formants produce relatively lesser error values. However, these positional errors are determined relative to Story's 1D vocal tract model, an oversimplified representation of the actual vocal tract. Hence, as future work, the 2.5D radiation model's acoustic characteristics need to be compared with a 3D vocal tract radiation model.

Additionally, our current 2.5D model should only be seen as simple approximations of the voice radiation mechanism. Nevertheless, the influence of few head details like the nose and cross-section of the mouth aperture is worth studying. As well as, our current 2.5D FDTD model does not include boundary losses for the tube depth. All these current limitations need to be incorporated for the future development of the 2.5D vocal tract acoustic model.

\section{Acknowledgements}
This work is supported by the Natural Sciences and Engineering Research Council (NSERC).

\eightpt
\printbibliography[heading=refs]

@article{takemoto2010acoustic,
  title={Acoustic analysis of the vocal tract during vowel production by finite-difference time-domain method},
  author={Takemoto, Hironori and Mokhtari, Parham and Kitamura, Tatsuya},
  journal={The Journal of the Acoustical Society of America},
  volume={128},
  number={6},
  pages={3724--3738},
  year={2010},
  publisher={ASA}
}

@article{vampola2015human,
  title={Human vocal tract resonances and the corresponding mode shapes investigated by three-dimensional finite-element modelling based on CT measurement},
  author={Vampola, Tom{\'a}{\v{s}} and Hor{\'a}{\v{c}}ek, Jarom\'{i}r and Laukkanen, Anne-Maria and {\v{S}}vec, Jan G},
  journal={Logopedics Phoniatrics Vocology},
  volume={40},
  number={1},
  pages={14--23},
  year={2015},
  publisher={Taylor \& Francis}
}

@article{arnela2014two,
  title={Two-dimensional vocal tracts with three-dimensional behavior in the numerical generation of vowels},
  author={Arnela, Marc and Guasch, Oriol},
  journal={The Journal of the Acoustical Society of America},
  volume={135},
  number={1},
  pages={369--379},
  year={2014},
  publisher={ASA}
}

@inproceedings{speed2009characteristics,
  title={Characteristics of two-dimensional finite difference techniques for vocal tract analysis and voice synthesis},
  author={Speed, Matt and Murphy, Damian and Howard, David M},
  booktitle={Tenth Annual Conference of the International Speech Communication Association},
  year={2009}
}

@article{yee1966numerical,
  title={Numerical solution of initial boundary value problems involving Maxwell's equations in isotropic media},
  author={Yee, Kane},
  journal={IEEE Transactions on antennas and propagation},
  volume={14},
  number={3},
  pages={302--307},
  year={1966},
  publisher={IEEE}
}

@inproceedings{zappi2016towards,
  title={Towards real-time two-dimensional wave propagation for articulatory speech synthesis},
  author={Zappi, Victor and Vasuvedan, Arvind and Allen, Andrew and Raghuvanshi, Nikunj and Fels, Sidney},
  booktitle={Proceedings of Meetings on Acoustics 171ASA},
  volume={26},
  number={1},
  pages={045005},
  year={2016},
  organization={ASA}
}

@article{mohapatra2019extended,
  title={An extended two-dimensional vocal tract model for fast acoustic simulation of single-axis symmetric three-dimensional tubes},
  author={Mohapatra, Debasish Ray and Zappi, Victor and Fels, Sidney},
  journal={arXiv preprint arXiv:1909.09585},
  year={2019}
}

@article{story2008comparison,
  title={Comparison of magnetic resonance imaging-based vocal tract area functions obtained from the same speaker in 1994 and 2002},
  author={Story, Brad H},
  journal={The Journal of the Acoustical Society of America},
  volume={123},
  number={1},
  pages={327--335},
  year={2008},
  publisher={ASA}
}

@phdthesis{thesis2015Arnela,
    title={Numerical production of vowels and diphthongs using finite element methods},
    author={Arnela Coll, Marc},
    school={Universitat Ramon Llull},
    year={2015}
}

@article{allen2015aerophones,
  title={Aerophones in flatland: Interactive wave simulation of wind instruments},
  author={Allen, Andrew and Raghuvanshi, Nikunj},
  journal={ACM Transactions on Graphics (TOG)},
  volume={34},
  number={4},
  pages={1--11},
  year={2015},
  publisher={ACM New York, NY, USA}
}

@article{yokota2002visualization,
  title={Visualization of sound propagation and scattering in rooms},
  author={Yokota, Takatoshi and Sakamoto, Shinichi and Tachibana, Hideki},
  journal={Acoustical science and technology},
  volume={23},
  number={1},
  pages={40--46},
  year={2002},
  publisher={ACOUSTICAL SOCIETY OF JAPAN}
}

@article{arnela2013effects,
  title={Effects of head geometry simplifications on acoustic radiation of vowel sounds based on time-domain finite-element simulations},
  author={Arnela, Marc and Guasch, Oriol and Al\'{i}as, Francesc},
  journal={The Journal of the Acoustical Society of America},
  volume={134},
  number={4},
  pages={2946--2954},
  year={2013},
  publisher={Acoustical Society of America}
}

@article{mohapatra2018limitations,
  title={Limitations Of Source-Filter Coupling In Phonation},
  author={Mohapatra, Debasish Ray and Fels, Sidney},
  journal={arXiv preprint arXiv:1811.07435},
  year={2018}
}

@article{sondhi1987hybrid,
  title={A hybrid time-frequency domain articulatory speech synthesizer},
  author={Sondhi, Man and Schroeter, Juergen},
  journal={IEEE Transactions on Acoustics, Speech, and Signal Processing},
  volume={35},
  number={7},
  pages={955--967},
  year={1987},
  publisher={IEEE}
}
\end{document}